\documentclass[apsmath, apssymb]{revtex4}
\usepackage{graphicx}
\begin{document}
\title{A Symmetry Scheme for Amino Acids Codons}
\author{J.Balakrishnan \thanks{E-mail : janaki@cmmacs.ernet.in }\\
CSIR Centre for Mathematical Modelling \& Computer Simulation, \\ 
Bangalore -- 560 037, India.}
\vspace{1.5cm}
\begin{flushright}
P.A.C.S. numbers ~~: 87.10.+e , ~82.39.Pj, ~02.90.+p
\end{flushright}
\vspace{1.5cm} 
\begin{abstract} 

Group theoretical concepts are invoked in a specific model to explain how 
only twenty amino acids occur in nature out of a possible sixty four. 
The methods we use enable us to justify the occurrence of the recently 
discovered twenty first amino acid selenocysteine, and also enables us to 
predict the possible existence of two more, as yet undiscovered amino acids. 
\end{abstract}
\maketitle
\newpage
\subsection*{1. ~Introduction}
The genetic code uses four ``letters'' or the bases adenine(A), thymine(T), 
guanine(G) and cytosine(C) in the four nucleotides constituting the DNA (or 
uracil(U) in the corresponding RNA template) by reading them in groups of three. 
 ~A and G are purine bases while C and T are pyrimidines. ~Like T, U pairs with 
A. ~During protein synthesis, these triplets of three bases 
(or codons) encode for specific amino acids. The genetic 
code however, is degenerate, and even though there are 64 possible codons, 
only 20 amino acids relevant to mammalian proteins actually occur in nature. 
It has remained intriguing that despite the redundancy of the codons, the 
genetic code did not expand any further and stopped at the number 20.  
It is therefore of interest to find out if the genetic code has any mathematical 
property which gets optimised when the number of codons becomes nearly thrice  
the number of the amino acids. 
We attempt here to answer this question by adapting some standard group 
theoretical methods of particle physics to molecular biology.    
The genetic code is nearly the same for all organisms --- non-canonical genetic 
codes are used in mitochondria and some protozoa~\cite{1}. Here, we consider 
only the universal genetic code.\\ 
Out of the 64 possible codons, it is now known that 61 code for the known 20  
amino acids --- the remaining three (UAG, UGA, and UAA) code for termination or 
``stop'' codons. The codon AUG for methionine also codes for the initiation of 
the translation process, and is therefore also called the ``start'' codon. It 
was discovered some years back that one of the stop codons, UGA, translates 
under certain circumstances to a twenty first amino acid 
selenocysteine~\cite{2}. It is certainly conceivable, that the other two stop 
codons  UAG and UAA  similarly code also for some as yet undiscovered amino 
acids.\\ 
Our approach is a semi-empirical one, but it enables us to not only justify the 
occurrence of selenocysteine, but it allows us also to predict the possible 
existence of two more, new, as yet undiscovered amino acids.  
We look at the hydrophilic and hydrophobic tendencies of the amino acid residues 
 constituting the proteins, as they play a key role in determining the 
conformation of a protein and the way it folds. \\  

The idea of using group theoretical techniques in studying the genetic code is 
not new --- references ~\cite{3} deal in length with searches for symmetries 
among Lie groups for trying to explain codon degeneracy in the genetic code. In 
their very interesting papers, the authors of ~\cite{3} view the universal 
genetic code as having evolved through a chain of successively broken symmetry 
events, from a primordial amino acid having a particular symmetry (which they 
assume to be Sp(6)). 
Their approach, however, does not presently account for the twenty first amino 
acid selenocysteine having properties similar to cysteine.\\
We approach the problem from a different point of view :  we have tried to show 
that it is possible to classify the 64 possible codons into well-defined 
multiplets --- the hydropathic properties of the amino acids they code for are 
determined by the multiplet they belong to.  Our approach has also a predictive 
power (presently lacking in the papers in ~\cite{3}), enabling one to 
approximately predict certain properties of two other possible, as yet 
undiscovered amino acids. \\
 
\subsection*{2. ~ The role of codons in protein synthesis} 
Protein synthesis is initiated by a process called transcription in which the 
cell makes a copy of the gene --- a messenger RNA (or mRNA) template, from the 
DNA with the help of an enzyme called RNA polymerase ~\cite{1,4}. 
Transcription stops when the enzyme reaches a ``stop'' sequence at the end of 
the gene, upon which the mRNA dissociates from the DNA, moves to the cytoplasm 
and gets attached at its Shine-Dalgarno sequence to a ribosomal RNA (rRNA) 
located within the ribosome. 
In the cytoplasm, transfer RNA (tRNA) molecules form complexes called 
aminoacyl-tRNAs with their respective amino acids, in a process driven by the 
enzymes aminoacyl-tRNA synthetases. An aminoacylated tRNA moves to the ribosome 
where its anticodon recognizes its corresponding complementary codon of the mRNA 
and incorporates the amino acid residue at the correct position specified by the 
mRNA codons into a growing peptide chain of the protein, this process being 
called translation. 
The folds governing the conformation of a protein are thus determined by its 
primary structure --- the sequence of the amino acids in the peptide chain,  
which in turn depend upon the sequence of codons in the functionally mature mRNA 
and the exon sequences in the DNA. 
\subsection*{3. ~Group theoretical methods for Codons} 
Keeping all of the above complex dynamics in mind, one could still try to look 
for any possible symmetries in the system, and see whether a much-simplified, 
minimal mathematical model for the codons could capture any of the physics and 
biochemistry of the actual biological system. \\
To begin with, we first recall the basic chemical structure of an amino acid. It 
has a central carbon called the $\alpha$-carbon which is attached to four groups 
--- a hydrogen atom, an acidic carboxyl (--COOH) group, a basic amino (--${\rm 
NH}_2$) group and a distinct side chain (--R) group --- this last group  
essentially determines its chemical properties.\\

The amino acids we know, can be classified into two broad categories on the 
basis of their solubilities in water : hydrophobic and hydrophilic. 
At pH 7.0, hydrophobic (non-polar) --R groups are contained by alanine, valine, 
leucine, isoleucine, proline, phenylalanine, tryptophan, methionine, cysteine 
and glycine.  
Hydrophilic side chains are polar, so that they can be further classified as 
acidic, basic or neutral, depending upon their charges --- at pH 7.0, lysine, 
arginine and histidine have basic side-chains, aspartate and glutamate are 
acidic and there are those which have polar but neutral side-chains:  
asparigine, glutamine, serine, threonine and tyrosine. 
Yet another consideration for classification of amino acids could be on the 
basis of whether the side chains are aliphatic or aromatic. \\

The categories or multiplets into which the amino acids fall, appear to reflect 
certain underlying internal symmetries. We know that while the base triplets 
(codons) do not constitute the amino acids, the base sequence within each codon 
dictates the identification of and translation to a particular amino acid. We 
can therefore hypothesise that the bases possess certain basic symmetries. \\

We look for the properties of the system which do not change or only 
approximately change in time, and the symmetries associated with the conserved 
quantities.\\
Let $T(a)$ represent a group of transformations which leave the Hamiltonian $H$ 
of the physical system invariant. We assume that the transformations are 
represented by unitary operators $U(a)$, (`$a$' denoting a parameter of the 
transformation) operating on a complex vector space~\cite{5}.  The eigenvalue 
equation for the system would be: 
\begin{equation} 
H\phi_n  =  E_n\phi_n
\end{equation}
where $\phi_n$ is an eigenfunction of $H$ with energy eigenvalue $E_n$. 
Operating on this with $U(a)$, one obtains :
\begin{eqnarray}
UH\phi_n &=& UHU^{-1}U\phi_n = E_nU\phi_n  \nonumber\\
{\rm or} ~~~ ~~~ H'{\phi'}_n  &=&  E_n{\phi'}_n
\end{eqnarray}
where we have let 
\begin{equation}
H' = UHU^{-1}  ~~~ ~~~ {\rm and} ~~~ ~~~ {\phi'}_n = U\phi_n ~~~  ~~ 
\end{equation}
Since $U$ leaves the Hamiltonian invariant : ~~ $H' = H$ , ~~ the state 
${\phi'}_n$ ~has the same energy as the state $\phi_n$. 
Operating on $\phi_n$ with $U(b)$, ~ `$b$'  being another parameter, would give 
another eigenstate of $H$ with the same energy $E_n$. 
The states which one obtains by operating with all $U$ on a given state can be 
expressed as linear combinations of a set of basis vectors spanning the subspace 
of eigenstates of H with a given energy. In general these vectors are the basis 
vectors of an irreducible representation and denote a set of states called a 
multiplet. All states of a multiplet are degenerate in the energy. \\

These basic techniques can be used to develop a symmetry scheme for the 
nucleotides and the codons. 
$A$ and $G$ can be regarded as different states of the same object, the purine, 
described by a state vector in an abstract, complex vector space, and similarly, 
$C$ and $T/U$  as different states of a pyrimidine. 
The purine and the pyrimidine state vectors are then each, two-component 
matrices :
$$\psi_R ~=~
\bordermatrix{~\cr
~&A\cr
~&G\cr}  ~~~~~~~{\rm and}~~~~~~~
\psi_Y ~=~
\bordermatrix{~\cr
~&C\cr
~&T\cr}  ~~{\rm or}
\bordermatrix{~\cr
~&C\cr
~&U\cr} 
$$

where $\psi_R$ and $\psi_Y$ denote the purine and pyrimidine state vectors 
respectively. A unitary transformation $U(\Lambda)$ which involves a 
rearrangement of the components, but which leaves the magnitude ${(\bar\psi_i 
\psi_i)}^{\frac{1}{2}}$, ~~($i = R,Y$) invariant can be written as: 
\begin{equation}
\psi_i' = U(\Lambda) \psi_i
\end{equation} 
Transition mutations involving replacement of one purine by another purine, or 
one pyrimidine by another pyrimidine can be represented by (4).  
The states representing A and G could be taken to correspond respectively to 
`up' and `down' states of $\psi_R$, with respect to a chosen axis in the 
internal vector space. A state intermediary between these two states could be 
regarded as a superposition of the two states, with the state having the larger 
probability measure, having the higher possibility of becoming the final state 
of the mutation. 
Proceeding similarly, we can define the full system of all the four bases by a 
four-component vector $\phi^i ~(i= 1,\dots,4)$ :  
\[ \vec \phi = \left ( \begin{array}{ll} 
A\\
G\\ 
C\\
U
\end{array} 
\right ) \]
A rotation through an angle $\vec \Lambda$ in this internal space which 
transforms  $\vec \phi$ to  $\vec \phi'$ :  
\begin{equation}
\vec \phi \rightarrow \vec \phi' = e^{i{\vec {\bf I}}\cdot{\vec {\bf 
\Lambda}}}\vec \phi
\end{equation}
where $I^k$ are  `k' number of ~4 $\times$ 4  matrices, and are representations 
of the generators of the transformation group, changes the state of the 
nucleotide system, but not the total number of nucleotides. Transversion 
mutations in which a purine is replaced by a pyrimidine, or vice-versa, are also 
covered by the transformation (5).\\

Since there are four different kinds of bases out of which 
three together code for one amino acid, we view the amino acids as arising out 
of ``3-base'' representations of the group SU(4). The purine and pyramidine 
bases are both hydrophobic and are turned inwards in the DNA structure.  The 
four bases are distinguished by three conserved numbers (which we denote by 
$J_1$, $J_2$ and $J_3$ (which reflects the rank 3 of the SU(4) group). This 
collection of four states serves as the basis vectors for a fundamental 
representation of SU(4).  A model with four bases could certainly correspond to 
a completely different symmetry group and not necessarily SU(4) --- we consider 
however, this possibility only. \\

The base triplets formed from the Kronecker product of three fundamental 
representations of SU(4), and arranged in the different multiplets (which are 
the decompositions of the product) represent at those positions, the amino acids 
they code for : 
\begin{equation}
\bf 4 \otimes 4 \otimes 4 =  \bf 20_S \oplus 20_M \oplus 20_M \oplus \bar 4
\end{equation}
where {\bf 4} is the fundamental representation of SU(4), $\bar {\bf 4}$ is the 
conjugate representation, and the subscripts {\bf S} and {\bf M} denote states 
which are formed from the symmetric combinations and the mixed symmetry 
combinations, respectively, of the product tensors. \\
Each multiplet is the realization of an irreducible representation of $SU(4)$, 
and because the members of each have masses which are not exactly, but only 
nearly degenerate, the $SU(4)$ symmetry is only an approximate symmetry. 
Notice that the total codon count of 64 is respected, but now the codons are 
grouped in separate multiplets. Each multiplet has a characteristic property 
which is shared by all its members. \\
Since the bases do not themselves constitute the amino acids, it follows that 
though the $J_i$ ($i = 1, \dots 3$),  are conserved numbers for the codons, they 
need not necessarily be additively conserved for the Kronecker product, since 
the permutations of the bases within a triplet only 
{\em code} to different amino acids.\\ 

We find that when we group together the amino acids into four categories:  
hydrophobic, weakly hydrophobic, hydrophilic and imino, and then try to adapt 
the SU(4) quantum numbers~\cite{6,7} for the quark triplets (baryons) to the 
codons, they fall beautifully into well-defined categories, as follows. \\

The numbers $J_i$ we have assigned for the base triplets are shown in Tables 
1--3 , within the standard format of the Universal Genetic Code. The alphabets 
within brackets in these tables are the conventional one-letter abbreviations 
for the various amino acids.  The left-hand column stands for the base present 
at the first (5') position of the codon, the top row denotes that in the second 
(middle) position of the codon, and the right-hand side column indicates the one 
present in the third (3') position of the codon. \\ 
The 3-dimensional plots of  $J_1$, $J_2$ and $J_3$  for all the codons are shown 
in Figures 1--4. 
Except for the case of proline (P), all the codons coding for a particular amino 
acid share the same $J_3$ number.\\

All the amino acids in Fig.1 are hydrophobic, and all in Fig.3 are polar and 
hydrophilic.  
It is of course well-known that there exist several different hydrophobicity 
scales, and there is no unique assignment of clear-cut hydrophobicity values for 
amino acids~\cite{8}. 
The amino acids in Fig.1, are, in general, widely accepted to be more 
hydrophobic than the others. 
We have classified proline (P) separately, as a realization of the conjugate 
${\bar 4}$  representation of SU(4), although it is very hydrophobic, since it 
is technically an imino (-- NH) acid rather than an amino (-- ${\rm NH_2}$) 
acid, as its side chain is bonded to the nitrogen as well as to the central 
$\alpha$ - carbon. \\

The amino acids cysteine (C),threonine (T), alanine (A), tryptophan (W), serine 
(S) and tyrosine (Y) in Fig.2 have been classified as weakly hydrophobic --- in 
fact some of them are slightly hydrophilic --- in some hydrophobicity scales, S 
and Y are categorized as being hydrophilic.\\
Notice that among the hydrophilic amino acids (hystidine (H), arginine (R), 
lysine (K), asparigine (N), glutamine (Q), aspartic acid (D), \& glutamic acid 
(E) ),  H, R and K are basic and occur with higher $J_3$ values than the acidic 
amino acids (D,E). \\ 
It is interesting to note that in our model, of the nine essential amino acids 
which are required for protein synthesis by adult humans, two (H and K) 
represent the hydrophilic multiplet, two (T and W ) represent the weakly 
hydrophobic multiplet, while all the amino acids except glycine (G)  (methionine 
(M), isoleucine(I), leucine (L), valine (V) and phenylalanine (F)) in the 
hydrophobic multiplet are essential.  

In general, the amino acids which have closer similarities between themselves, 
occur nearer to each other within each multiplet. The amino acids in each 
multiplet are also in close conformity with the standard suggested amino acid 
substitutions based on the Dayhoff matrix~\cite{9}  wherein amino acid residues 
which are near to each other in the Dayhoff plot are good candidates for mutual 
exchange for conservative mutations in proteins.\\

It is extremely interesting that with the assignment of the $J_i$  numbers as in 
Tables 1--3, the codon UGA which usually codes for ``Stop'', falls in the 
multiplet of weakly hydrophobic amino acids, and has the same $J_3$ value of 2 
as cysteine (C).  UGA codes also for the newly discovered twenty first amino 
acid selenocysteine (SeC) --- the sulfur atom in C is replaced by selenium in 
SeC.\\  

On the basis of these observations, one could similarly predict the existence of 
two more as yet undiscovered amino acids --- the codons UAG and UAA which are 
presently known to code only for the ``Stop'' signal.  
One could hypothesise that if UAG were to code for a new (twenty second) amino 
acid, then that would have properties similar to H, and similarly, UAA if coded 
for a twenty third amino acid would have properties similar to K or R, even 
though these two codons, both differ from Y only at the wobble position. \\

In our symmetry scheme for codons, we have not yet found it possible to assign 
$J_i$  numbers for each base individually so as to give additively, the total 
$J_i$  numbers for each base triplet coding for the amino acids, in a consistent 
way. This is not unreasonable, since as emphasised before, the bases {\em do 
not} constitute the amino acids.\\
Synonymous codons occur at different frequencies even though they all code for 
the same amino acid. Correspondingly, the tRNA molecule for the codon used more, 
occurs in larger amounts than its isoacceptors.  
This fact is reflected in the differing $J_1$ and $J_2$ values for synonymous 
codons. The probability for the occurrence of a particular synonymous codon 
would be weighted by $J_i$ dependent factors in the corresponding partition 
function and its free energy.\\
Thus, the occurrence of only twenty one amino acids out of a possible sixty 
four, can be explained in a satisfactory manner within our scheme.
\subsection*{Discussion} 
We have classified the 64 codons within a semi-empirical model which very 
closely resembles the decomposition of the Kronecker product  $\bf 4 \otimes 4 
\otimes 4$  of SU(4), after assuming that bases $A$, $C$, $G$ and $T/U$ can be 
regarded as different states of a vector in an abstract, complex vector space.   
We represent transition and transversion mutations involving replacement of 
purines and pyramidines, by rotations through an angle in this internal space. 
Our model explains the existence of synonymous codons (thus explaining how 
twenty one amino acids (including selenocysteine) have been found to occur so 
far out of a possible sixty four). It also enables us to predict the possible 
existence of two more, as yet undiscovered amino acids.\\
The stability of a fully folded native protein structure is a consequence of a 
balance between the hydropathies of the constituent amino acid residues in its 
primary structure, electrostatic interactions and hydrogen bonding. It would 
thus be a useful exercise to incorporate the ideas in this paper in an 
analytical manner to approach the protein folding problem, since incorporation 
of all the internal symmetries into the partition function is essential to get 
the correct form of the free energy of the system. \\
Our ideas and methods could also be very useful in providing a rigorous 
mathematical basis for studying DNA replication and protein synthesis using  
quantum algorithms. 

\subsection*{Acknowledgement} 
I have benefited from very helpful discussions with Dr.Preeti Chhabra, Dept. of 
Cell Biology, School of Medicine, University of Virginia, Charlottesville, which 
I gratefully acknowledge. I am very grateful to Dr.D.J.Toms, Dept. of Physics, 
University of Newcastle-upon-Tyne, U.K., and  B.Ashok, Dept. of Physics, 
University of Massachussetts, Amherst, for going through the manuscript and for 
very helpful comments. I would also like to express my gratitude to the referee 
for bringing to my notice the papers listed in ref.[3]. 

\begin{tabular}{|l|r|r|r|r|rl|}
\hline
\multicolumn{7}{|c|}{~~~~Table 1:~~$J_1$:}\\
\multicolumn{7}{|c|}{Second~~~~~~~~Position}\\
\hline
 &U &C &A &G & &	\\
\hline
 &-1/2 (F)      &-1/2 (S)        &0 (Y)       &-1/2 (C)        &U & \\
U &1/2 (F)       & 1/2 (S)         &0 (Y)       & 1/2 (C)        &C & \\
 &-1/2 (L)      & 1/2 (S)        &0 (Stop)       &0 (SeC)       &A & \\
 &1/2 (L)      &-1/2 (S)        &0 (Stop)        &0 (W)        &G & \\ 
\hline
 & & & & & & \\
P &-1 (L)      &-1/2 (P)       &-1/2 (H)        &-1 (R)        &U &P\\
o &0 (L)       &0   (P)        &1/2 (H)         &0 (R)        &C &o\\
s C&0 (L)       &0   (P)       &-1/2 (Q)         &0 (R)        &A &s \\
i &1 (L)       &1/2 (P)       & 1/2 (Q)         &1 (R)        &G &i \\
t & & & & & &t\\
\hline
i &-1/2 (I)      & -1  (T)     &  -1/2 (N)       &-1 (S)        &U &i\\
o &0 (I)        &0  (T)        &1/2 (N)        &1 (S)        &C &o\\
n A &1/2 (I)       & 0  (T)      & -1/2 (K)     &-1/2 (R)       &A &n\\
 &0 (M)        &1  (T)        &1/2 (K)      &1/2 (R)        &G &\\
1 & & & & & &3\\
\hline
 &-3/2 (V)      &-1/2 (A)         &-1 (D)        &-1 (G)       &U & \\
 &-1/2 (V)      & 1/2 (A)         & 1 (D)        & 0 (G)       &C & \\
G &1/2 (V)       &1/2 (A)          &0 (E)         &1 (G)        &A & \\
 &3/2 (V)      &-1/2 (A)          &0 (E)         &0 (G)        &G & \\
\hline
\end{tabular}
\newpage
\begin{tabular}{|l|r|r|r|r|rl|}
\hline
\multicolumn{7}{|c|}{~~~~Table 2:~~$J_2$:}\\
\multicolumn{7}{|c|}{Second~~~~~~~~Position}\\
\hline
 &U &C &A &G & &   \\
\hline
 &-1 (F)     &-1 (S)        &0 (Y)          &1 (C)          &U &\\ 
 &-1 (F)      &1 (S)        &0 (Y)          &1 (C)          &C &\\
U &0 (L)      &-1 (S)       &-1 (Stop)       &0 (SeC)        &A &\\
 &0 (L)       &1 (S)        &0 (Stop)      &-1 (W)          &G &\\
\hline
 & & & & & &\\
P &1 (L)       &0 (P)        &1 (H)          &1 (R)          &U &P\\
o &-1 (L)      &0 (P)        &1 (H)          &1 (R)          &C &o\\
s C&1 (L)       &1 (P)        &1 (Q)          &1 (R)          &A &s\\
i&1 (L)       &0 (P)        &1 (Q)          &1 (R)          &G &i\\
t& & & & & &t\\
\hline
i &1 (I)       &1 (T)       &-1 (N)          &0 (S)          &U &i\\
o A &0 (I)       &1 (T)       &-1 (N)          &0 (S)          &C &o\\
n &1 (I)       &1 (T)        &0 (K)          &0 (R)          &A &n\\
 &1 (M)       &1 (T)        &0 (K)          &0 (R)          &G &\\
1& & & & & &3\\
\hline
&1 (V)       &0 (A)        &0 (D)          &0 (G)          &U &\\
&1 (V)       &0 (A)        &0 (D)          &0 (G)          &C &\\
G &1 (V)       &0 (A)        &0 (E)          &0 (G)          &A &\\
&1 (V)       &0 (A)        &0 (E)         &-2 (G)          &G &\\
\hline
\end{tabular} 
\newpage
\begin{tabular}{|l|r|r|r|r|rl|}
\hline
\multicolumn{7}{|c|}{~~~~Table 3:~~$J_3$:}\\
\multicolumn{7}{|c|}{Second~~~~~~~~Position}\\
\hline
 &U &C &A &G & &   \\
\hline
 &0 (F)      &0 (S)        &0 (Y)           &2 (C)          &U & \\
 &0 (F)      &0 (S)        &0 (Y)           &2 (C)          &C & \\
U &1 (L)      &0 (S)        &1 (Stop)        &2 (SeC)        &A & \\ 
 &1 (L)      &0 (S)        &2 (Stop)        &1 (W)          &G & \\
\hline
 & & & & & & \\
P &1(L)      &1 (P)        &2 (H)           &1 (R)          &U &P\\
o &1 (L)      &0 (P)        &2 (H)           &1 (R)          &C &o\\
s C &1 (L)      &1 (P)        &0 (Q)           &1 (R)          &A &s \\
i &1 (L)      &1 (P)        &0 (Q)           &1 (R)          &G &i \\
t & & & & & &t\\
\hline
i &2 (I)      &1 (T)        &0 (N)           &0 (S)          &U &i\\
o &2 (I)      &1 (T)        &0 (N)           &0 (S)          &C &o\\
n A &2 (I)      &1 (T)        &1 (K)           &1 (R)          &A &n\\
 &3 (M)      &1 (T)        &1 (K)           &1 (R)          &G &\\
1 & & & & & &3\\
\hline
 &0 (V)      &1 (A)        &0 (D)           &0 (G)          &U & \\
 &0 (V)      &1 (A)        &0 (D)           &0 (G)          &C & \\
G &0 (V)      &1 (A)        &0 (E)           &0 (G)          &A & \\
 &0 (V)      &1 (A)        &0 (E)           &0 (G)          &G & \\
\hline
\end{tabular}
\newpage
{\bf Figure Captions}\\
\vspace*{1.5cm}\\
{\bf Fig.1.} {Hydrophobic Amino Acids ~~~ ($20_{\rm S}$ of SU(4))}\\
\vspace{0.2in}\\
{\bf Fig.2.} {Weakly Hydrophobic Amino Acids ~~~ ($20_{\rm M}$ of SU(4))}\\
\vspace{0.2in}\\
{\bf Fig.3.} {Hydrophilic Amino Acids ~~~ ($20_{\rm M}$ of SU(4))}\\ 
\vspace{0.2in}\\
{\bf Fig.4.} {Imino (Proline) ~~~ (${\bar 4}$ of SU(4))}\\
\newpage 
{\includegraphics[width=6.5cm]{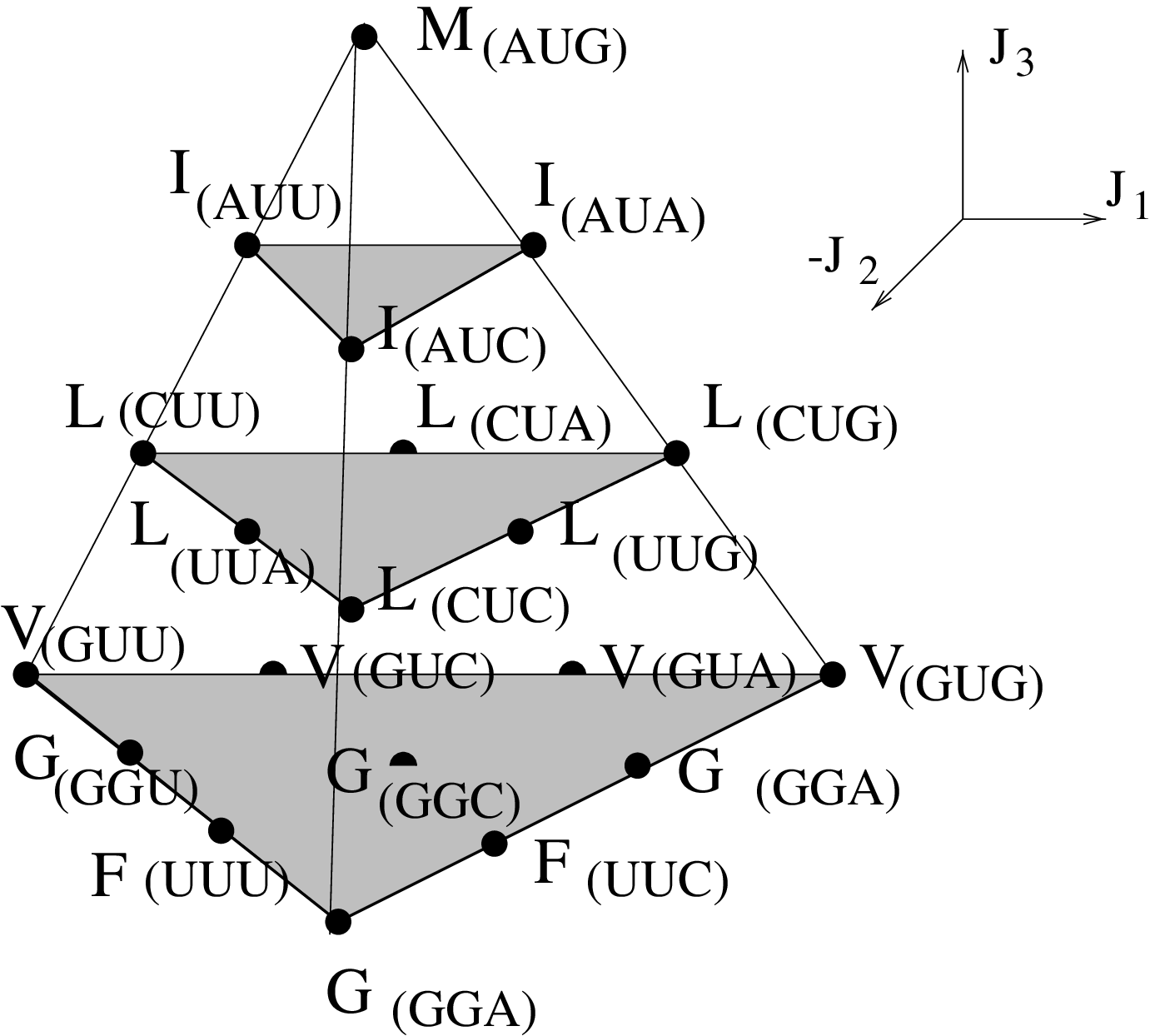}}
\vskip 2in
{\bf Fig.1.~~~Hydrophobic Amino Acids ~~~ ($20_{\rm S}$ of SU(4))}\\
\vspace*{.5in}\\ 
\hspace*{1.5in} J.Balakrishnan  \hspace{0.7in}Phys.Rev.E\\
\newpage

{\includegraphics[width=6.5cm]{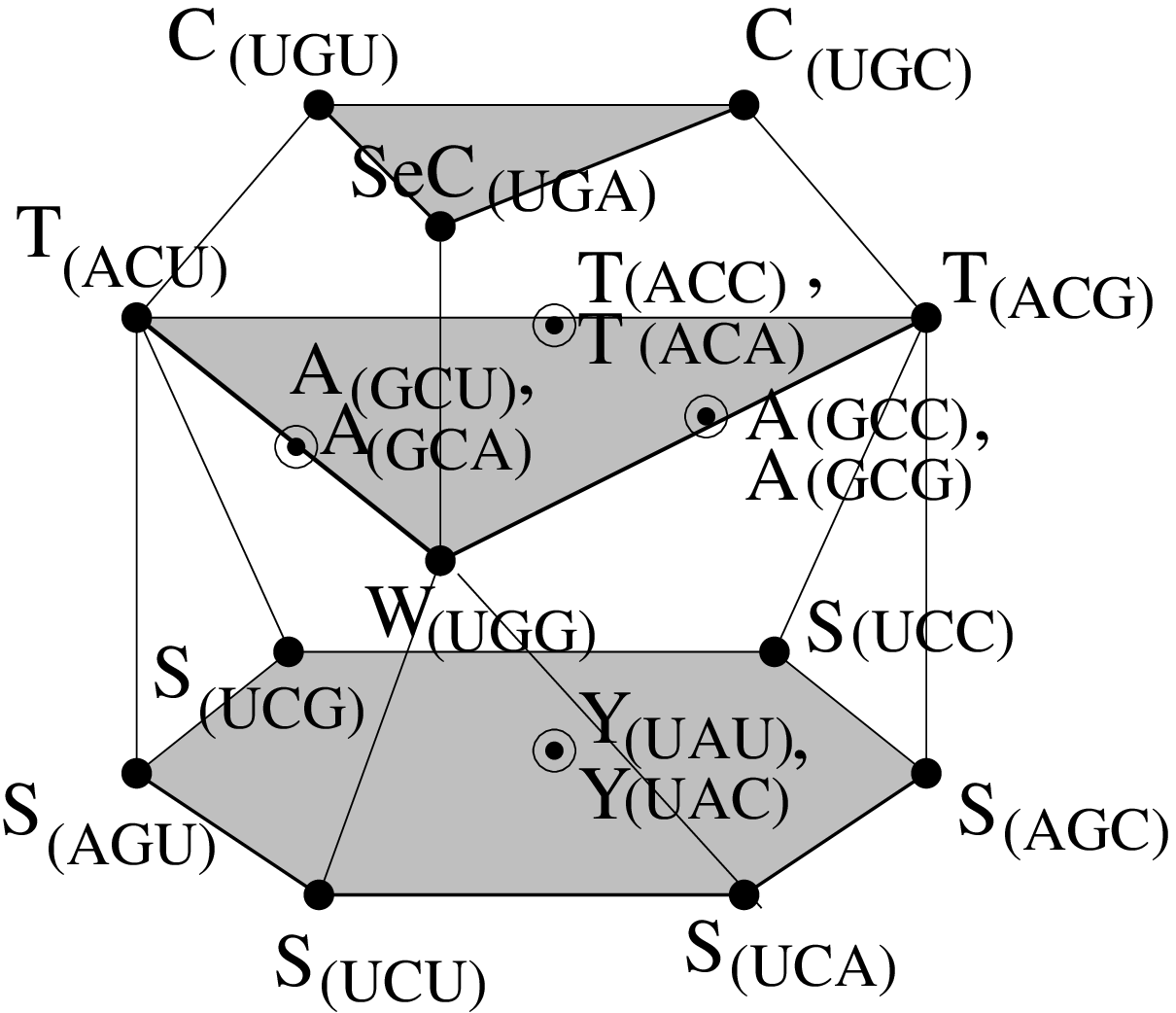}
\vskip 2in
{\bf Fig.2.~~~Weakly Hydrophobic Amino Acids ~~~ ($20_{\rm M}$ of SU(4))}
\vspace*{.5in}\\
\hspace*{1.5in} J.Balakrishnan  \hspace{0.7in}Phys.Rev.E\\
\newpage

{\includegraphics[width=6.5cm]{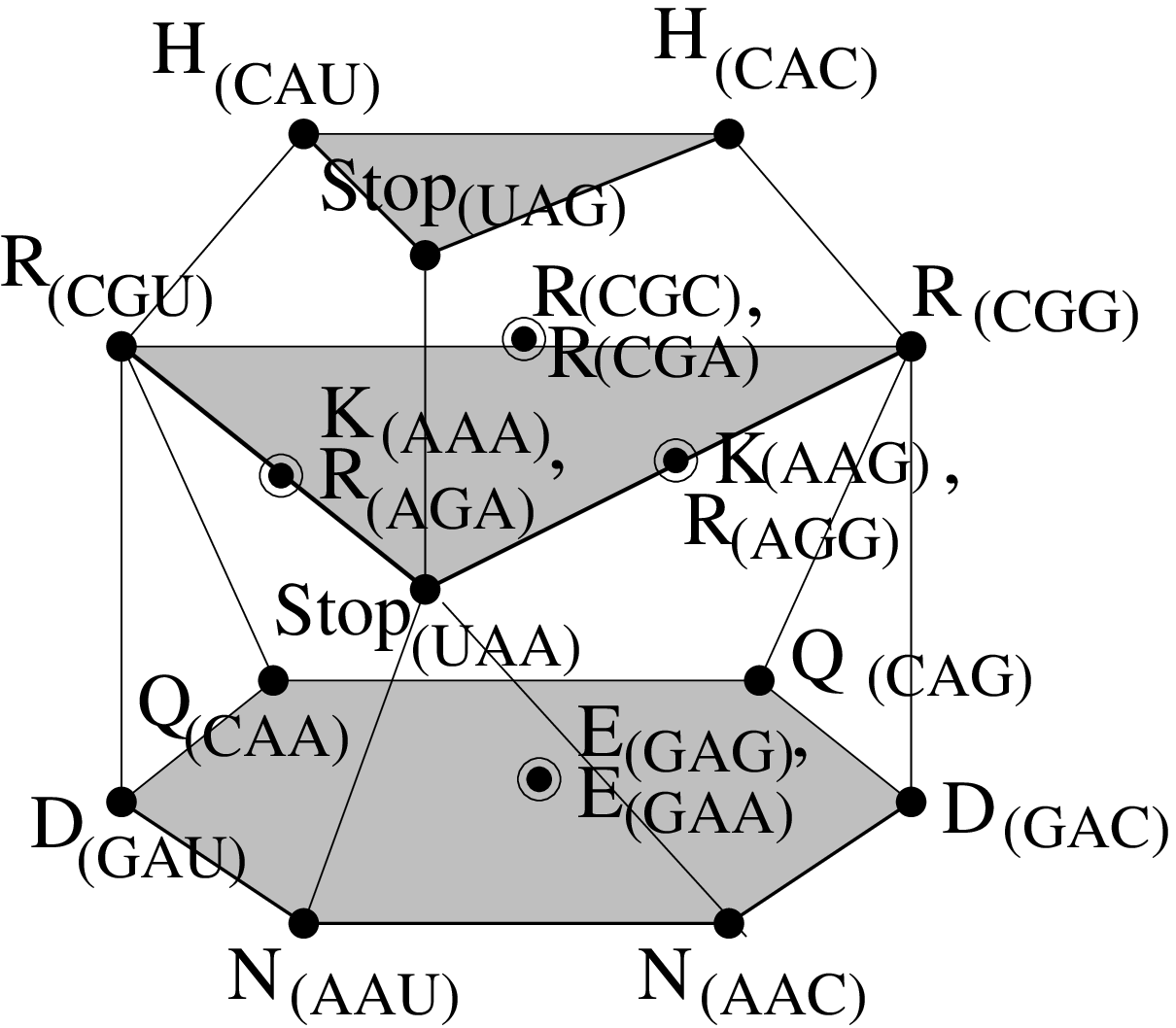}}
\vskip 2in
{\bf Fig.3.~~~Hydrophilic Amino Acids ~~~ ($20_{\rm M}$ of SU(4))}
\vspace*{.5in}\\
\hspace*{1.5in} J.Balakrishnan  \hspace{0.7in}Phys.Rev.E\\
\newpage

{\includegraphics[height=5cm,width=6cm]{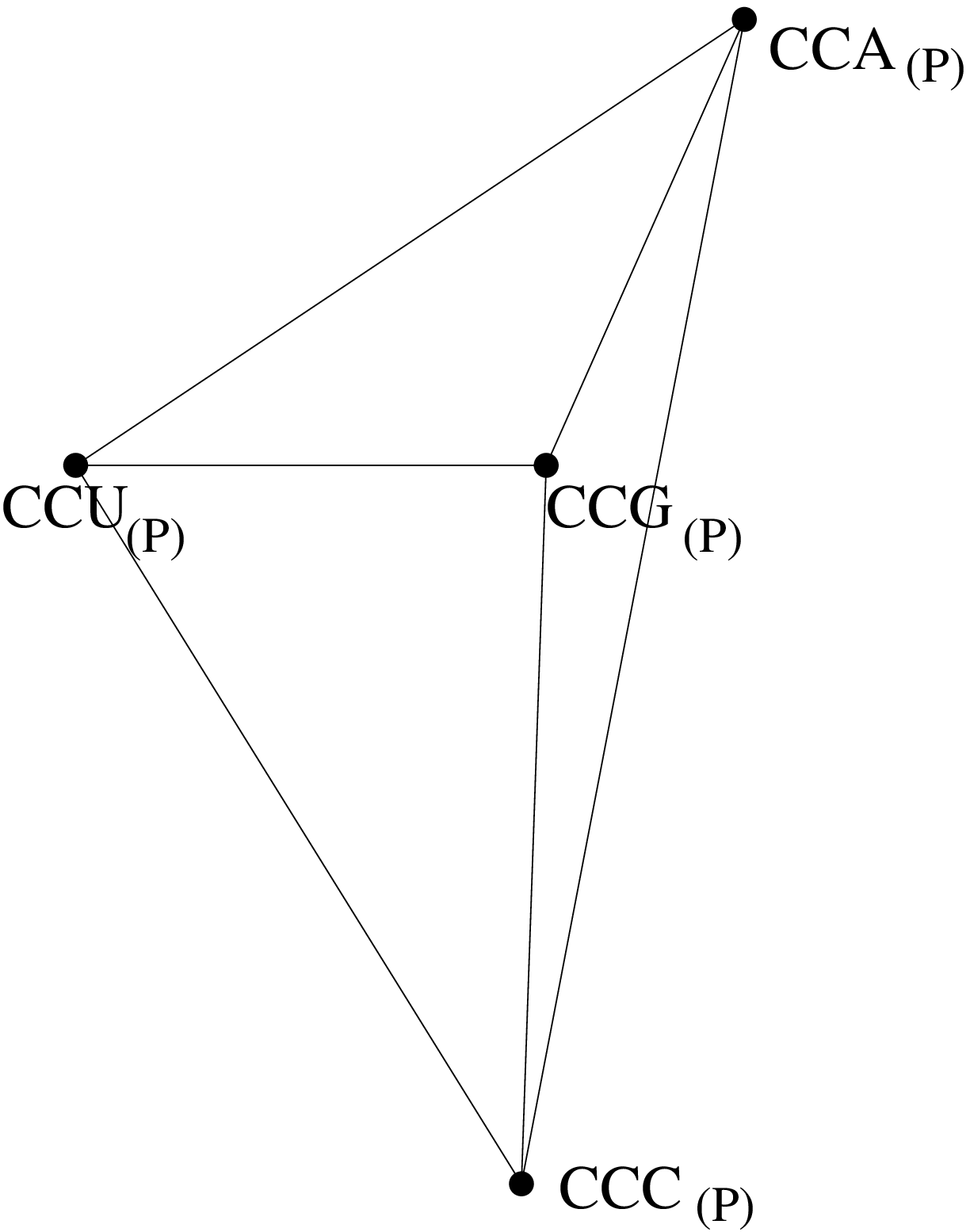}}
\vskip 2in
\hspace*{1in}{\bf Fig.4.~~~Imino (Proline) ~~~ (${\bar 4}$ of SU(4))}
\vspace*{.5in}\\
\hspace*{1.2in} J.Balakrishnan  \hspace{0.7in}Phys.Rev.E\\
\end{document}